\documentstyle[epsbox]{article}

\def\abstract#1{\vskip 7mm 
        \begin{center}{\large Abstract}\par \smallskip
                \begin{minipage}[c]{12cm}
                        \small #1
                \end{minipage}
        \end{center}
}
\def\title#1{\begin{center}{\Large\bf #1}\end{center}}
\def\author#1{\vskip 5mm \begin{center}{#1}\end{center}}
\def\address#1{\begin{center}{\it #1}\end{center}}
\makeatletter
\@ifundefined{lesssim}{}{}
\@ifundefined{gtrsim}{}{}
\def\vereq#1#2{\lower3pt\vbox{\baselineskip1.5pt \lineskip1.5pt
\ialign{$\m@th#1\hfill##\hfil$\crcr#2\crcr\sim\crcr}}}
\makeatother

\begin{document}

\title{%
Aspects of black hole entropy
}
\author{%
Shinji Mukohyama~\footnote{E-mail:mukoyama@uvic.ca}
}
\address{%
Department of Physics and Astronomy, 
University of Victoria\\ 
Victoria, BC, Canada V8W 3P6\\
and\\
Canadian Institute for Theoretical Astrophysics, 
University of Toronto\\
Toronto, ON, M5S 3H8
}
\abstract{
There have been many attempts to understand the statistical origin of
black-hole entropy. Among them, entanglement entropy and the brick
wall model are strong candidates. In this paper, first, we show that 
the entanglement approach reduces to the brick wall model when we seek
the maximal entanglement entropy. After that, the stability of the
brick wall model is analyzed in a rotating background. It is shown
that in the Kerr background without horizon but with an inner boundary
a scalar field has complex-frequency modes and that, however, the
imaginary part of the complex frequency can be small enough compared
with the Hawking temperature if the inner boundary is sufficiently
close to the horizon, say at a proper altitude of Planck scale. 
Hence, the brick wall model is well defined even in a rotating
background if the inner boundary is sufficiently close to the
horizon. These results strongly suggest that the entanglement
approach is also well defined in a rotating background. 
}


\section{Introduction}

Black hole entropy is given by a mysterious formula called the
Bekenstein-Hawking formula~\cite{Bekenstein,Hawking}: 
$S_{BH}=A/4l_{pl}^2$, 
where $A$ is area of the horizon. There have been many attempts to
understand the statistical origin of the black-hole
entropy~\cite{D-thesis}.

Entanglement entropy~\cite{BKLS,Srednicki} is one of the strongest
candidates of the origin of black hole entropy. It is originated from
a direct-sum structure of a Hilbert space of a quantum system: for an
element $|\psi\rangle$ of the Hilbert space ${\cal F}$ of the form 
%
\begin{equation}
 {\cal F} = {\cal F}_I \bar{\otimes} {\cal F}_{II},
	\label{eqn:F=F1*F2}
\end{equation}
the entanglement entropy $S_{ent}$ is defined by 
%
\begin{equation}
 S_{ent} = -{\bf Tr}_I[\rho_I\ln\rho_I],\quad
 \rho_I  = {\bf Tr}_{II}|\psi\rangle\langle\psi|.
\end{equation}
Here $\bar{\otimes}$ denotes a tensor product followed by a suitable
completion and ${\bf Tr}_{I,II}$ denotes a partial trace over 
${\cal F}_{I,II}$, respectively.

On the other hand, there is another strong candidate for the origin of 
black hole entropy: the brick wall model introduced by
'tHooft~\cite{tHooft}. In this model, thermal atmosphere in
equilibrium with a black hole is considered. In this situation, we
encounter with two kinds of divergences in physical quantities. The
first is due to infinite volume of the system and the second is due
to infinite blue shift near the horizon. We are not interested in the
first since it represents contribution from matter in the far
distance. Hence we introduce an outer boundary in order to make our
system finite. It is the second divergence that we would like to
associate with black hole entropy. Namely, it can be shown by
introducing a Planck scale cutoff that entropy of the thermal
atmosphere near the horizon is proportional to the area of the horizon
in Planck units.

In this paper, first, we show that the brick wall model seeks the
maximal value of the entanglement entropy. In other words,
the entanglement approach reduces to the brick wall model when we
seek the maximal value of the entanglement entropy. 
After that, we analyze the stability of the brick wall model in a
rotating background. We show that the time scale of the ergoregion
instability is much longer than the relaxation time scale of the
thermal state with the Hawking temperature. In the latter time scale
ambient fields should settle in the thermal state. Thus, the brick
wall model is well defined even in a rotating background. These
results strongly suggest that the entanglement approach is also well
defined in a rotating background.

This paper is organized as Follows. In Sec.~\ref{sec:BW-ent} the
relation between the entanglement entropy and the brick wall model is
shown. In Sec.~\ref{sec:stability-BW} the stability of the brick wall 
model is analyzed in the Kerr background. Sec.~\ref{sec:summary} is
devoted to a summary of this paper.


\section{Brick wall entropy as maximal entanglement entropy}
	\label{sec:BW-ent}

\begin{figure}
 \begin{center}
  \epsfile{file=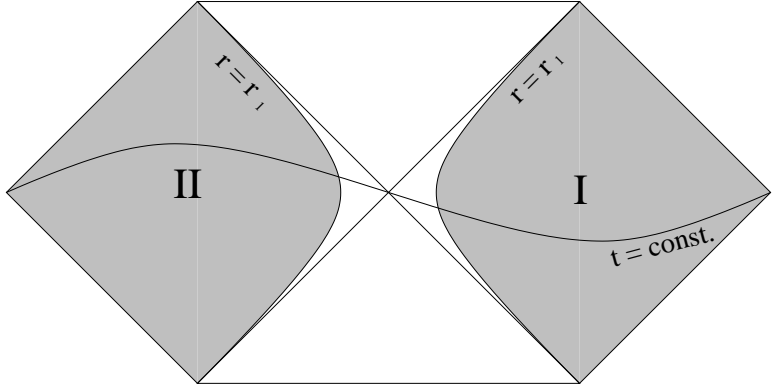,scale=0.5}
 \end{center}
\caption{
The Kruskal-like extension of the static, spherically
symmetric black-hole spacetime. We consider only the regions
satisfying $r>r_1$ (the shaded regions $I$ and $II$ ).
}
\label{fig:Kruskal}
\end{figure}

For simplicity, we consider a minimally coupled, real scalar field
described by the action 
%
\begin{equation}
 S = -\frac{1}{2}\int d^4x\sqrt{-g}\left[
        g^{\mu\nu}\partial_{\mu}\phi\partial_{\nu}\phi
        + \mu^2\phi^2\right], 
\end{equation}
in a spherically symmetric, static black-hole space-time
%
\begin{equation}
 ds^2 = -f(r)dt^2 + \frac{dr^2}{f(r)} + r^2d\Omega^2.
\end{equation}
We denote the area radius of the horizon by $r_0$ and the surface
gravity by $\kappa_0$ ($\ne 0$):
%
\begin{eqnarray}
 f(r_0) = 0,\quad
 \kappa_0 = \frac{1}{2}f'(r_0).
\end{eqnarray}
We quantize the system of the scalar field with respect to the Killing 
time $t$ in a Kruskal-like extension of the black hole spacetime. 
The corresponding ground state is called the Boulware state and its 
energy density is known to diverge near the horizon. 
Although we shall only consider states with bounded energy density, it
is convenient to express these states as excited states above the
Boulware ground state for technical reasons. 
Hence, we would like to introduce an ultraviolet cutoff $\alpha$ with 
dimension of length to control the divergence. 
The cutoff parameter $\alpha$ is implemented so that we only consider
two regions satisfying $r>r_1$ (shaded regions $I$ and $II$ in 
{\it Figure}~\ref{fig:Kruskal}), where $r_1$ ($>r_0$) is determined by 
%
\begin{equation}
 \alpha = \int_{r_0}^{r_1}\frac{dr}{\sqrt{f(r)}}. 
\end{equation}
[Evidently, the limit $\alpha\to 0$ corresponds to the limit 
$r_1\to r_0$. Thus, in this limit, the whole region in which 
$\partial /\partial t$ is timelike is covered.]
Strictly speaking, we also have to introduce outer boundaries, say at 
$r=L$ ($\gg r_0$), to control the infinite volume of the constant-$t$
surface. However, even if there are outer boundaries, the
following arguments still hold.

In this situation, there is a natural choice for division of the
system of the scalar field: let ${\cal H}_I$ be the space of mode
functions with supports in the region $I$ and ${\cal H}_{II}$ be the 
space of mode functions with supports in the region $II$. 
Thence, the space ${\cal F}$ of all states are of the form
(\ref{eqn:F=F1*F2}), where ${\cal F}_I$ and ${\cal F}_{II}$ are
defined as symmetric Fock spaces constructed from ${\cal H}_I$ and
${\cal H}_{II}$, respectively:
%
\begin{equation}
 {\cal F}_{I,II} \equiv 
    \mbox{\boldmath C} \oplus {\cal H}_{I,II} \oplus 
    \left({\cal H}_{I,II}\bar{\otimes} {\cal H}_{I,II}\right)_{sym}
    \oplus \cdots.
\end{equation}
Here $(\cdots)_{sym}$ denotes the symmetrization.

Let us investigate what kind of condition should be imposed for our
arguments to be self-consistent. 
A clear condition is that the backreaction of the scalar field to the
background geometry should be finite.
For the brick wall model this condition is satisfied. Namely, in
Ref.~\cite{Mukohyama&Israel}, it was shown that the total mass of the
thermal atmosphere of quantum fields is actually bounded. 
Thus, also for our system, we would like to impose the condition that 
the contribution $\Delta M$ of the subsystem ${\cal F}_I$ to the mass
of the geometry should be bounded in the limit 
$\alpha\to 0$.

It is easily shown that $\Delta M$ is given by
%
\begin{equation}
 \Delta M \equiv 
	-\int_{x\in I}T^t_t 4\pi r^2dr = H_I,
\end{equation}
where $H_I$ is the Hamiltonian of the subsystem ${\cal F}_I$ with
respect to the Killing time $t$.
Hence, the expectation value of $\Delta M$ with respect to a state
$|\psi\rangle$ of the scalar field is decomposed into the contribution 
of excitations and the contribution from the zero-point energy: 
%
\begin{equation}
 \langle\psi |\Delta M|\psi\rangle = E_{ent} + \Delta M_{B},
\end{equation}
where $E_{ent}$ is entanglement energy defined by
%
\begin{equation}
 E_{ent} \equiv \langle\psi |:H_I:|\psi\rangle, \label{eqn:Eent}
\end{equation}
and $\Delta M_{B}$ is the zero-point energy of the Boulware state. 
Here, the colons denote the usual normal ordering. 
[This definition of entanglement energy corresponds to
$E_{ent}^{(I')}$ in Ref.~\cite{MSK1998} and $\langle :H_2:\rangle$ in
Ref.~\cite{D-thesis}.]

Since the Boulware energy $\Delta M_{B}$ diverges as 
$\Delta M_{B}\sim -AT_H\alpha^{-2}$ in the limit 
$\alpha\to 0$~\cite{Mukohyama&Israel}, we should impose the condition 
%
\begin{equation}
 E_{ent} \simeq |\Delta M_B|, \label{eqn:SBC}
\end{equation}
where $A=4\pi r_0^2$ is the area of the horizon, $T_H=\kappa_0/2\pi$
is the Hawking temperature. We call this condition {\it the small
backreaction condition (SBC)}. Note that the right hand side of SBC
(\ref{eqn:SBC}) is independent of the state $|\psi\rangle$.

Now, we shall show that the Hartle-Hawking state is a maximum of the 
entanglement entropy in the space of quantum states satisfying SBC. 
For this purpose, we have a more general statement for a quantum
system with a state-space of the form (\ref{eqn:F=F1*F2}): 
{\it a state of the form 
%
\begin{equation}
 |\psi\rangle = {\cal N} \sum_n 
	e^{-E_n/2T}|n\rangle_{I}\otimes|n\rangle_{II}
	\label{eqn:HH-state}
\end{equation}
is a maximum of the entanglement entropy in the space of states with
fixed expectation value of the operator $E_{I}$ defined by 
%
\begin{equation}
 E_{I} = \left(\sum_n E_n|n\rangle_{I}\cdot{}_{I}\langle n|\right)
	\otimes 
	\left(\sum_m |m\rangle_{II}\cdot{}_{II}\langle m|\right),
\end{equation}
provided that the real constant $T$ is determined so that the
expectation value of $E_I$ is actually the fixed value. Here,
$\{|n\rangle_I\}$ and $\{|n\rangle_{II}\}$ ($n=1,2,\cdots$) are bases
of the subspaces ${\cal F}_I$ and ${\cal F}_{II}$, respectively, and
$E_n$ are assumed to be real and non-negative.}
(For a proof of this statement, see Ref.~\cite{Mukohyama9904}.)
Note that this statement is almost the same as the following statement
in statistical mechanics: a canonical state is a maximum of
statistical entropy in the space of states with fixed energy, provided
that the temperature of the canonical state is determined so that the
energy is actually the fixed value. 

Note that the expectation value of $E_I$ is equal to the entanglement
energy (\ref{eqn:Eent}), providing that $|n\rangle_I$ and $E_n$ are an
eigenstate and an eigenvalue of the normal-ordered Hamiltonian $:H_I:$
of the subsystem ${\cal F}_I$. 
Hence, for the system of the scalar field, the above general statement
insists that the state (\ref{eqn:HH-state}) is a maximum of the
entanglement entropy in the space of states satisfying SBC, which
corresponds to fixing the entanglement entropy. 
Off course, in this case, the constant $T$ should be determined so
that SBC (\ref{eqn:SBC}) is satisfied. 
The value of $T$ is easily determined as $T=T_H$ by using the
well-known fact that the negative divergence in the Boulware energy
density can be canceled by thermal excitations if and only if
temperature with respect to the time $t$ is equal to the Hawking
temperature.

Finally, we obtain the statement that the Hartle-Hawking
state~\cite{Hartle&Hawking} is a maximum of entanglement entropy in
the space of quantum states satisfying SBC since the Hartle-Hawking
state is actually of the form (\ref{eqn:HH-state}) with
$T=T_H$~\cite{Israel1976}.
[Strictly speaking, in order to obtain the Hartle-Hawking state, we
have to take the limit $\alpha\to 0$ (and $L\to\infty$). However, the
following arguments still hold for a finite value of $\alpha$ (and
$L$).] 
The corresponding reduced density matrix is the thermal state with
temperature equal to the Hawking temperature. Therefore, the maximal
entanglement entropy is equal to the thermal entropy with the Hawking
temperature, which is sought in the brick wall model.

In summary the brick wall model seeks the maximal value of
entanglement entropy~\cite{Mukohyama9904,Mukohyama9907}. 
In other words, the entanglement approach reduces to the brick wall
model when we seek the maximal entanglement entropy.


\section{Stability of the brick wall model in a rotating background}
	\label{sec:stability-BW}

Originally, the brick wall model was proposed in a spherically
symmetric, static background, say, the Schwarzschild
background. Hence, it seems interesting to see how this model is
extended to a rotating background, say, the Kerr
background~\cite{Lee&Kim,HKPS,Ho&Kang,Frolov&Fursaev}. 
However, it is known that in a rapidly rotating spacetime without
horizon a field has complex-frequency
modes~\cite{Sato&Maeda,Comins&Schutz} and that there is the so called
ergoregion instability~\cite{Kang}. Thus, it might be expected that
the brick-wall model in rotating background might be unstable and
unsuitable for the origin of black hole entropy.

For simplicity, let us consider the Kerr spacetime as a
background. The metric is given by 
%
\begin{equation}
 ds^2 = -\left(1-\frac{2Mr}{\Sigma}\right)dt^2 
	+ \frac{\Sigma}{\Delta}dr^2 + \Sigma d\theta^2
	+ R^2\sin^2\theta d\varphi^2
	- \frac{4Mar}{\Sigma}\sin^2\theta d\varphi dt,
\end{equation}
where
%
\begin{eqnarray}
 \Sigma & = & r^2 + a^2\cos^2\theta,\nonumber\\
 \Delta & = & r^2 + a^2 - 2Mr,\nonumber\\
 \Sigma R^2 & = & (r^2 + a^2)^2 - a^2\Delta\sin^2\theta.
\end{eqnarray}

We only consider the region $r\geq r_1$ in this spacetime: we impose
the Dirichlet boundary condition on the field $\phi$ at
$r=r_1$. (For more general boundary conditions, see
Ref.~\cite{Solodukhin}.) Hereafter we assume that
$r_1>M+\sqrt{M^2-a^2}$: there is no horizon in the region 
$r\geq r_1$.

\begin{figure}
 \begin{center}
  \epsfile{file=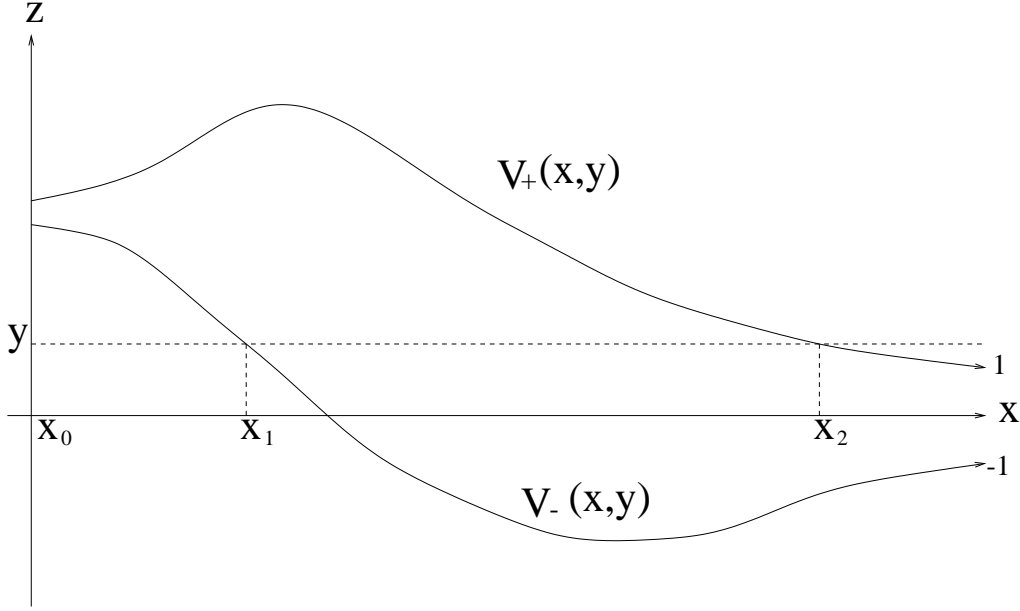,scale=0.5}
 \end{center}
\caption{
The typical form of the graphs $z=V_{\pm}(x,y)$ is written on a fixed
$y$ plane. Note that $V_{\pm}$ depend on $y$ as well as $x$ through
the eigen value $\lambda$ of equation (\ref{eqn:eq-S}). 
However, asymptotic behavior of $V_{\pm}$ in the limit $x\to\pm\infty$
does not depend on $y$. 
}
\label{fig:potential}
\end{figure}

It is well known that in this background the field equation of a
massless real (Klein-Gordon) scalar field becomes separable. In fact,
we can find a solution of the field equation of the form
%
\begin{equation}
 f = \frac{u(r)}{\sqrt{r^2+a^2}}S(\theta),
\end{equation}
where $S$ satisfies
%
\begin{equation}
 \frac{1}{\sin\theta}\frac{d}{d\theta}
	\left(\sin\theta\frac{dS}{d\theta}\right)
	+ \left[\lambda - a^2\mu^2(y^2 - 1)\sin^2\theta
		-\frac{m^2}{\sin^2\theta}\right]S = 0,
	\label{eqn:eq-S}
\end{equation}
and the equation for $u$ can be written as 
%
\begin{equation}
 \frac{d^2u}{dx^2} + (y - V_+)(y - V_-) u =0
	\label{eqn:d2u/dx2}
\end{equation}
by introducing the non-dimensional tortoise coordinate $x$ by 
%
\begin{equation}
 \mu^{-1}\frac{dx}{dr} = \frac{r^2+a^2}{\Delta}, 
\end{equation}
or
%
\begin{equation}
 \mu^{-1}x = r + \frac{M}{\sqrt{M^2-a^2}}\left[
	r_+ \ln\left(\frac{r-r_+}{r_+-r_-}\right)
	- r_- \ln\left(\frac{r-r_-}{r_+-r_-}\right)\right]. 
\end{equation}
Here, $\mu$ is the mass of the scalar field,
$r_{\pm}=M\pm\sqrt{M^2-a^2}$, $y\equiv\omega/\mu$ and $V_{\pm}$ are
defined by 
%
\begin{eqnarray}
 \mu V_{\pm} & = &\frac{2maMr}{(r^2+a^2)^2} \pm
	\sqrt{\frac{\Delta\tilde{\mu}^2}{r^2+a^2}},\nonumber\\
 \tilde{\mu}^2 & = & \mu^2+\frac{\lambda}{(r^2+a^2)}
	-\frac{m^2a^2(r^2+a^2+2Mr)}{(r^2+a^2)^3}
	+\frac{2Mr+a^2}{(r^2+a^2)^2}
	-\frac{6Ma^2r}{(r^2+a^2)^3}. 
\end{eqnarray}
Note that in the horizon limit $r\to r_+$ (or $x\to -\infty$) both of
$V_{\pm}$ approach to the same value
%
\begin{equation}
 V_{\pm} \to \frac{ma}{2\mu Mr_+}, \label{eqn:limit-V}
\end{equation}
and that $V_{\pm}$ approaches to $\pm 1$, respectively, in the limit
$r\to\infty$ (or $x\to\infty$). (See Fig.~\ref{fig:potential}.)

In Ref.~\cite{Mukohyama9910}, the existence of complex frequency modes 
was shown by examining a scattering amplitude for real-frequency
waves. The strategy is based on the following expected form of the
scattering amplitude $S$ near a pole $y=y_R+iy_I$, providing that
$|y_R|\gg |y_I|$~\cite{Comins&Schutz}. 
%
\begin{equation}
 S = e^{2i\delta_0}\times\frac{y-y_R+iy_I}{y-y_R-iy_I},
\end{equation}
where $\delta_0$ is a constant phase. Hence, if we can obtain this
form of a scattering amplitude by analyzing real-frequency waves then
we find an outgoing normal mode corresponding to $y=y_R+iy_I$ and an
incoming normal mode corresponding to $y=y_R-iy_I$. After that we
should investigate whether these normal modes converge or diverge in
the limit of $x\to\infty$. If these normal modes converge then they
give complex-frequency mode functions.

By analyzing Eq.~(\ref{eqn:d2u/dx2}) in the WKB approximation and
using the above strategy, we can obtain the following
results~\cite{Mukohyama9910}. 
\begin{itemize}
\item [(a)] When $ma>2\mu Mr_+$ and $1<y<V_-(x_1)$, we obtain a set of
complex-frequency modes. However, the imaginary part $\Im\omega$ of
the complex frequency is small enough:
%
\begin{equation}
 T_{BH}^{-1}\Im\omega \sim 
	\pm \pi e^{-2\eta}
	\left|\ln\left(\frac{r_1-r_+}{r_+-r_-}\right)\right|^{-1}
	\to 0 \quad ( r_1\to r_+ ),
\end{equation}
where $T_{BH}$ is the Hawking temperature of the Kerr background. 
\item [(b)] When $ma<-2\mu Mr_+$ and $V_+(x_1)<y<-1$, we obtain a set
of complex-frequency modes. However, the imaginary part $\Im\omega$ of
the complex frequency is small enough: 
%
\begin{equation}
 T_{BH}^{-1}\Im\omega \sim 
	\pm \pi e^{-2\eta}
	\left|\ln\left(\frac{r_1-r_+}{r_+-r_-}\right)\right|^{-1}
	\to 0 \quad ( r_1\to r_+ ).
\end{equation}
\item[(c)] In other cases, there is no complex-frequency modes. 
\end{itemize}

In this section we have shown that in the Kerr background without
horizon but with an inner boundary a scalar field has
complex-frequency modes and that the imaginary part of the complex
frequency is small enough compared with the Hawking temperature if the 
inner boundary is sufficiently close to the horizon, say at a proper
altitude of Planck scale. Hence, the time scale of the ergoregion
instability is much longer than the relaxation time scale of the
thermal state with the Hawking temperature. In the latter time scale
ambient fields should settle in the thermal state. In this sense
ergoregion instability is not so catastrophic. Thus, the brick wall
model is well defined if the inner boundary is sufficiently close to
the horizon.


\section{Summary and discussion}
	\label{sec:summary}

In this paper we have analyzed the entanglement entropy and the brick
wall model. In Sec.~\ref{sec:BW-ent} we have shown that the
entanglement approach reduces to the brick wall model when we seek
the maximal value of the entanglement entropy. In
Sec.~\ref{sec:stability-BW} we have shown that the brick wall model is 
well defined even in a rotating background. These two results strongly 
suggest that the entanglement approach is also well defined in a
rotating background.


\end{document}